\documentclass{ssdbm}
\usepackage{amssymb}
\usepackage{subfigure}
\usepackage{graphicx}
\usepackage{color}
\usepackage{url}
\usepackage[symbol*]{footmisc}

\newcommand{\para}[1]{\noindent{\bf #1:}}

\conferenceinfo{SSDBM}{'13, July 29 - 31 2013, Baltimore, MD, USA}
\global\boilerplate={Permission to make digital or hard copies of all or part of this work for personal or classroom use is granted without fee provided that copies are not made or distributed for profit or commercial advantage and that copies bear this notice and the full citation on the first page. Copyrights for components of this work owned by others than ACM must be honored. Abstracting with credit is permitted. To copy otherwise, or republish, to post on servers or to redistribute to lists, requires prior specific permission and/or a fee. Request permissions from Permissions@acm.org.}
\global\copyrightetc{Copyright 2013 ACM 978-1-4503-1921-8/13/07 \$15.00}

\title{The Open Connectome Project Data Cluster: Scalable Analysis and Vision for High-Throughput Neuroscience}

\numberofauthors{20}
\author{
\alignauthor Randal Burns\footnotemark[1]
\alignauthor William Gray Roncal\footnotemark[2]
\alignauthor Dean Kleissas\footnotemark[2] 
\and
\alignauthor Kunal Lillaney\footnotemark[1] 
\alignauthor Priya Manavalan\footnotemark[1] 
\alignauthor Eric Perlman\footnotemark[4] 
\and
\alignauthor Daniel R. Berger\footnotemark[6]
\alignauthor Davi D. Bock\footnotemark[4]
\alignauthor Kwanghun Chung\footnotemark[9]
\and
\alignauthor Logan Grosenick\footnotemark[9]
\alignauthor Narayanan Kasthuri\footnotemark[5]
\alignauthor Nicholas C. Weiler\footnotemark[10] 
\and
\alignauthor Karl Deisseroth\footnotemark[9]
\alignauthor Michael Kazhdan\footnotemark[3] 
\alignauthor Jeff Lichtman\footnotemark[5]
\and
\alignauthor R. Clay Reid\footnotemark[7]
\alignauthor Stephen J. Smith\footnotemark[10]
\alignauthor Alexander S. Szalay\footnotemark[8]
\and
\alignauthor Joshua T. Vogelstein\footnotemark[3] 
\alignauthor R. Jacob Vogelstein\footnotemark[2] 
}
\additionalauthors{
{\footnotemark[1]~Department of Computer Science and the Institute for Data Intensive Engineering and Science, Johns Hopkins University}\\
{\footnotemark[2]~Johns Hopkins University Applied Physics Laboratory}\\
{\footnotemark[3]~Department of Statistical Science and Mathematics and the Institute for Brain Science, Duke University}\\
{\footnotemark[4]~Janelia Farm Research Campus, Howard Hughes Medical Institute}\\
{\footnotemark[5]~Department of Molecular and Cellular Biology, Harvard University}\\
{\footnotemark[6]~Department of Computational Neuroscience, Massachusetts Institute of Technology}\\
{\footnotemark[7]~Allen Institute for Brain Science}\\
{\footnotemark[8]~Department of Physics and Astronomy and the Institute for Data Intensive Engineering and Science, Johns Hopkins University}\\
{\footnotemark[9]~Department of Bioengineering, Stanford University}\\
{\footnotemark[10]~Department of Molecular and Cellular Physiology, Stanford University}\\
}

\date{}
       
\begin{document}

\maketitle


\begin{abstract}
We describe a scalable database cluster for the spatial analysis and annotation of high-throughput brain imaging data, initially for 3-d electron microscopy image stacks, but for time-series and multi-channel data as well. The system was designed primarily for workloads that build {\em connectomes}---neural connectivity maps of the brain---using the parallel execution of computer vision algorithms on high-performance compute clusters. These services and open-science data sets are publicly available at \url{openconnecto.me}.

The system design inherits much from NoSQL scale-out and data-intensive computing architectures. We distribute data to cluster nodes by partitioning a spatial index. We direct I/O to different systems---reads to parallel disk arrays and writes to solid-state storage---to avoid I/O interference and maximize throughput. All programming interfaces are RESTful Web services, which are simple and stateless, improving scalability and usability. We include a performance evaluation of the production system, highlighting the effectiveness of spatial data organization.
\end{abstract}



%
%

\footnotesize{This work was supported by the National Institutes of Health (NIBIB 1RO1EB016411-01) and the National Science Foundation (OCI-1244820 and CCF-0937810).

\section{Introduction}


The neuroscience community faces a scalability crisis as new high-throughput imaging technologies
come online.  Most notably, electron microscopes that image serial sections now produce data 
at more than one terabyte per day.  Data at this scale can no longer be stored, 
managed, and analyzed on workstations in the labs of the scientists that collect them. 

In response to this crisis, we have developed a community database cluster based on 
the principles of data-intensive computing~\cite{Gray06} and Open Science~\cite{nielsen11}.  
Labs contribute imaging data to the Open Connectome Project (OCP).  In exchange, OCP provides 
storage and analysis Web-services that relieve the data management burden for neuroscientists 
and create an incentive to make data publicly available.  (Data analysis products may be kept 
private.)  This open science model replicates that pioneered by the Sloan Digital Sky Survey~\cite{sdss}
for observational astronomy, democratizing world-class data sets by making them freely available
over the Internet.
To date, the model has been well received by leading neuroscientists; OCP 
manages the largest image stack \cite{bock11}
and the most detailed neural reconstruction \cite{kasthuri11} collected to date and have partnered with both teams
to receive data streams from the next generation of instruments.

Data-intensive computing will create the capability to reconstruct neural circuits at a scale that is
relevant to characterizing brain systems and will help to solve grand challenge problems, such as
building biologically-inspired computer architectures for machine learning and discovering 
``connecto-pathies''---signatures for brain disease based on neural connectivity that are 
diagnostic and prognostic.  Previous studies have been limited by analysis capabilities to image
volumes representing tens of neurons \cite{bock11}.  At present, automated reconstruction tools are 
neither sufficiently accurate nor scalable~\cite{Reina2011,Jain2010}.  The largest dense neural 
reconstruction describes only 100s of GB of data \cite{Reina2011a}.
As a consequence, current analyses depends on humans using manual annotation tools to describe 
neural connectivity \cite{TrakEM2} or to correct the output of computer vision algorithms~\cite{eyewire}.   
The gap between the  size of the system and the state of current technology
scopes the problem.  A graph representing the human brain has a fundamental size of 
$10^{11}$ vertices (neurons) and $10^{15}$ edges (synapses).  Mouse brains have $10^9$ nodes 
and $10^{13}$ vertices.  
We conclude that manual annotation cannot reach these scales.
More accurate computer vision and scalable data systems must be developed in order to realize 
the ultimate goal of a full reconstruction of the brain---the human connectome or human brain map.

The Open Connectome Project was specifically designed to be a scalable data infrastructure for 
parallel computer vision algorithms that discover neural connectivity through image processing.
Researchers have many different approaches to the task.  For example, the segmentation 
process, which divides the image into bounded regions, may be performed with feed-forward 
neural networks \cite{Jain2011} or by geometric tracing  \cite{Kaynig2010}.
However, neural reconstruction has some fundamental properties that
we capture in system design.  Algorithms realize parallelism through a geometric decomposition 
of the data.  OCP provides a {\em cutout} service to extract spatially contiguous regions of the
data, including projections to lower dimensions.   We use indexes derived from space-filling curves
to partition data which makes cutout queries efficient and (mostly) uniform across lower dimensional 
projections.  Partitions are striped across disk arrays to realize parallel I/O and to
multiple nodes for scale-out.   Vision algorithms 
output descriptions of neural connectivity.  We capture these in a relational database of neural object 
metadata~\cite{cajal3d} linked to a spatial {\em annotation} database that stores the structures.  
Algorithms run in multiple phases, assembling structure from the output of previous stages, 
e.g.~fusing previous segmentations into neurons.  We support queries across images and annotations.
Annotation databases are spatially registered to images.  
Finally, we support spatial queries for individual objects and regions that are used in analysis to
extract volumes, find nearest neighbors, and compute distances.

The Open Connectome Project stores more than 50 unique data sets
totaling more than 75TB of data.  Connectomes range from the macro (magnetic resonance imaging of 
human subjects at 1 mm$^3$) to the micro (electron microscopy of mouse visual cortex at
4nm $\times$ 4nm $\times$ 40nm).  We have demonstrated scalable computer vision in the system by 
extracting more than 19 million synapse detections from a 4 trillion pixel image volume: one quarter scale of
the largest published EM connectome data~\cite{bock11}.  This involved a cluster of three physical nodes
with 186 cores running for three days, communicating with the OCP cutout and annotation service over the Internet.

\begin{figure}
\begin{center}
\includegraphics[width=3in]{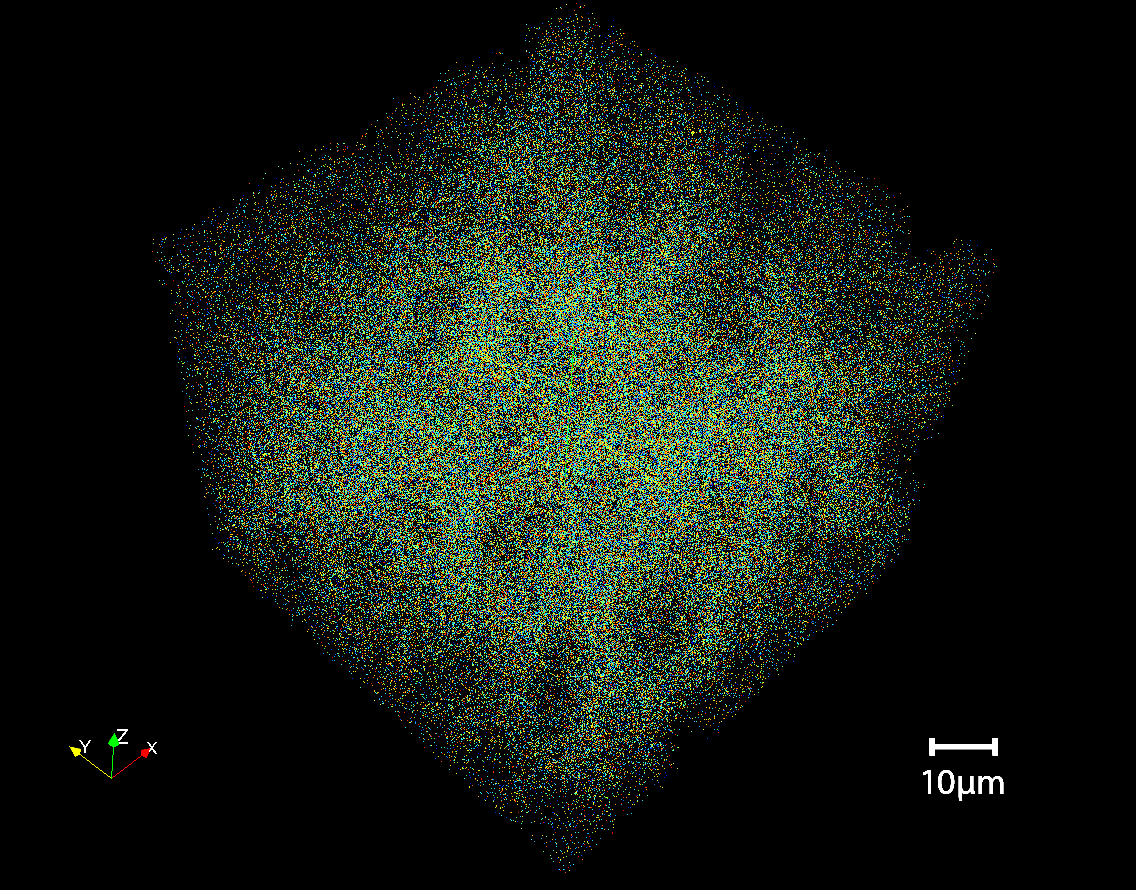}
\caption{Visualization of the spatial distribution of synapses detected in the mouse visual 
cortex of Bock et al. \cite{bock11}.}
\label{fig:bock11synapses}
\end{center}
\end{figure}

\begin{figure}[ht]
\begin{center}
\includegraphics[width=3in]{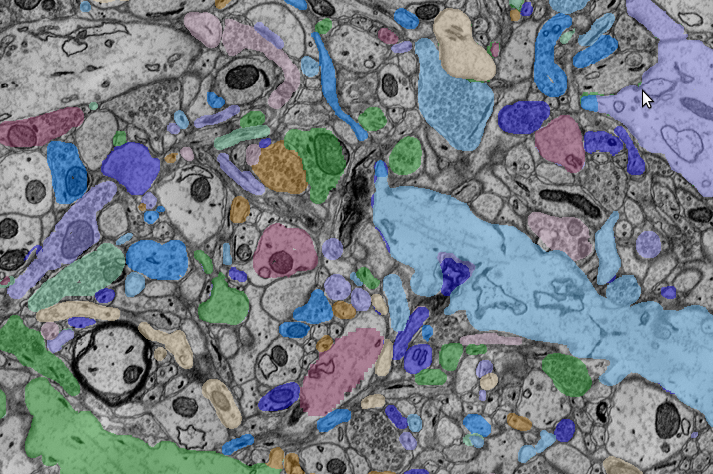}
\caption{Electron microscopy imaging of a mouse somatosensory cortex \cite{kasthuri11} overlaid by 
  manual annotations describing neural objects, including axons, dendrites, and synapses.  These images
  were cutout from two spatially registered databases and displayed in the CATMAID Web viewer \cite{catmaid}.}
\label{fig:kat11data}
\end{center}
\end{figure}

\section{Data and Applications}
\label{sec:data}

We present two examples data sets and the corresponding analysis as use cases for
Open Connectome Project services.  The data themselves are quite similar: high-resolution 
electron microscopy of a mouse brain.  However, the analyses of these data 
highlight different services.

The {\tt bock11} data \cite{bock11} demonstrates state-of-the-art scalability and the use of 
parallel processing to perform computer vision.  The image data are the largest published 
collection of high-resolution images, covering a volume of roughly 450x350x50 microns with 20 trillion voxels 
at a resolution of 4x4x40nm. 
Volumes at this scale just start to contain the connections between neurons.  
Neurons have large spatial extent and connections can be analyzed when both cells 
and all of the connection wiring (dendrite/synapse/axon) lie within the volume.   
We are using this data to explore the spatial distribution of synapses, identifying
clusters and outliers to generate a statistical model of where neurons connect.
Our synapse-finding vision algorithm extracts
more than 19 millions locations in the volume (Figure \ref{fig:bock11synapses}).
We have not yet characterized the precision and recall of this technique.
Thus, this exercise is notable for its scale only; we ran 20 parallel instances and processed 
the entire volume in less than 3 days.  For comparison, Bock et al. \cite{bock11} collected
this data so that they could manually trace 10 neurons, 245 synapses, and 185 postsynaptic 
targets over the course of 270 human days.
We build a framework for running this task within the  
LONI \cite{loni} parallel execution environment.

The {\tt kasthuri11} data \cite{kasthuri11} shows how spatial analysis can be performed using 
object metadata and annotations (Figure \ref{fig:kat11data}).  
This data has the most detailed and accurate manual annotations.
Two regions of 1000x1000x100 and 1024x1024x256 voxels have been densely reconstructed, 
labeling every structure in the volume.  
Three dendrites that span the entire 12000x12000x1850 voxel volume have had all synapses that attach 
to dendritic spines annotated. 
OCP has ingested all of these manual annotations, including 
object metadata for all structures and a spatial database of annotated regions.
This database has been used to answer questions about the spatial distribution of 
synapses (connections) with respect to the target dendrite (major neuron branch).
This analysis proceeds based on: (1) using metadata to get the identifiers of all 
synapses that connect to the specified dendrite and then (2) querying the spatial extent
of the synapses and dendrite to compute distances.  The latter stage can be done by 
extracting each object individually or specifying a list of objects and a region and
having the database filter out all other annotations.
We also use the densely annotated regions as a ``ground truth'' for evaluating machine 
vision reconstruction algorithms.

\section{Data Model}

The basic storage structure in OCP is a dense multi-dimensional spatial array partitioned
into cuboids (rectangular subregions) in all dimensions.  Cuboids in OCP are
similar in design and goal to chunks in ArrayStore \cite{Soroush11}.   Each cuboid gets assigned an index using 
a Morton-order space-filling curve (Figure \ref{fig:zorderpart}).  Space-filling curves organize data 
recursively so that any power-of-two aligned subregion is wholly contiguous in the index 
\cite{Perlman07}.  
Space-filling curves minimize the number of discontiguous regions needed to retrieve a
convex shape in a spatial database \cite{moon01}.  While the Hilbert curve has the best 
properties in this regard, we choose the Morton-order curve for two reasons.  It is 
simple to evaluate using bit interleaving of offsets in each dimension, unlike other curves 
that are defined recursively.  Also, cube addresses are strictly non-decreasing in each 
dimension so that the index works on subspaces.  Non-decreasing offsets also aid in interpolation, filtering, 
and other image processing operations~\cite{Kanov12,Crow84}.

Image data contain up to 5 dimensions and often exhibit anisotropy.  For example, serial section electron microscopy
data come from imaging sections created by slicing or ablating a sample.
The resolution of image plane (XY) is determined by the instrument and the sections (Z) by the sectioning technique.
The difference in resolution is often a factor of 10.
The fourth and fifth dimension arise in other imaging modalities.  Some techniques produce time-series,
such as functional magnetic-resonance and light-microscopy of calcium channels.  
Others image multiple channels that 
correspond to different proteins or receptors (Figure \ref{fig:clarity}).
These data are studied by correlating multiple channels, e.g.~spatial co-occurence or exclusion, 
to reveal biology.

\begin{figure}
\begin{center}
\includegraphics[width=3in]{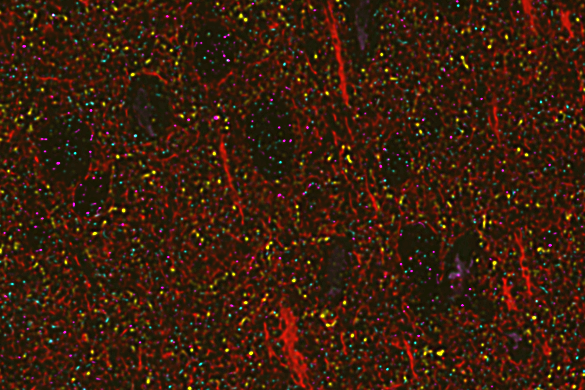}
\caption{Visualization of six channels array tomography data courtesy of Nick Weiler and Stephen Smith
\cite{Orourke12, Micheva10}.
Data were drawn from a 17-channel database and rendered by the OCP cutout service.}
\label{fig:clarity}
\end{center}
\end{figure}

\subsection{Physical Design}

We store a multi-resolution hierarchy for each image data set, so that analyses and visualization can choose
the appropriate scale.  For EM data, each lower resolution reduces the data size by a factor of four, 
halving the scale in X and Y.  Typically, we do not scale Z, because it is poorly resolved.  We also do 
not scale time or channels.  The {\tt bock11} data has nine levels and {\tt kasthuri11} six.  As an example of choosing 
scale, our {\tt bock11} synapse detector runs on high resolution data, because synapses have limited spatial extent 
(tens of voxels in any dimension) .  We detect synapses at resolution one: four times smaller and four times
faster than the raw image data.  We found that the algorithms was no less accurate at this scale.
We analyze large structures that cannot contain synapses, such as blood vessels and cell bodies, to mask out 
false positives.  We developed this software for this analysis, but many of the techniques follow those 
describe in ilastik \cite{ilastik}.  We conduct the analysis at resolution $5$ in which each voxel represents a 
(32,32,1) voxel region in the raw data.  The structures are large and detectable at low resolution and the computation
requires all data to be in memory to run efficiently.

\begin{figure}
\begin{center}
\includegraphics[width=3.2in]{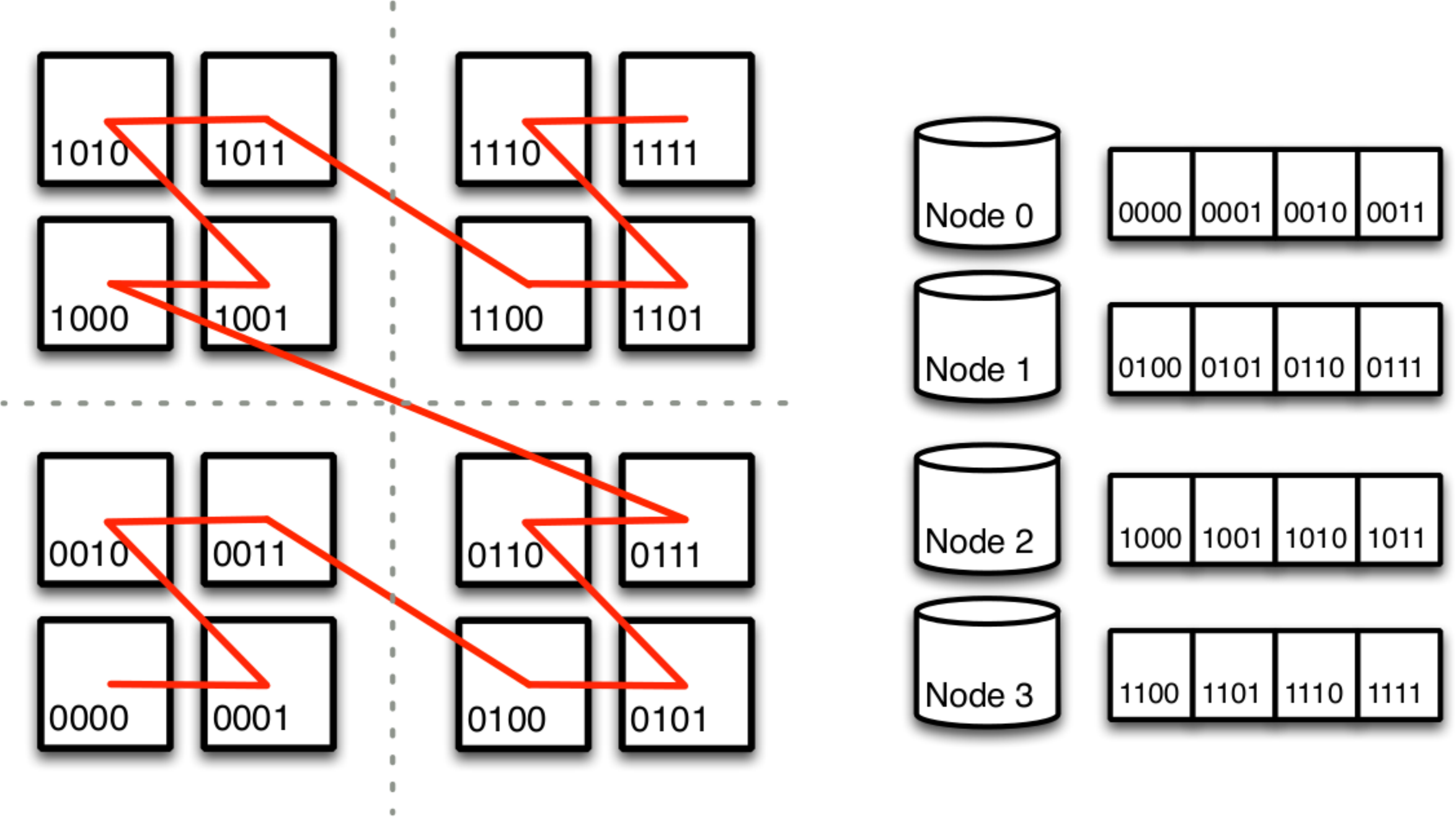}
\caption{Partitioning the Morton (z-order) space-filling curve.  For clarity, the figure shows
16 cuboids in 2-dimensions mapping to four nodes.  The z-order curve is recursively defined and scales in 
dimensions and data size.}
\label{fig:zorderpart}
\end{center}
\end{figure}

\begin{figure}[t]
\begin{center}
\includegraphics[width=3.2in]{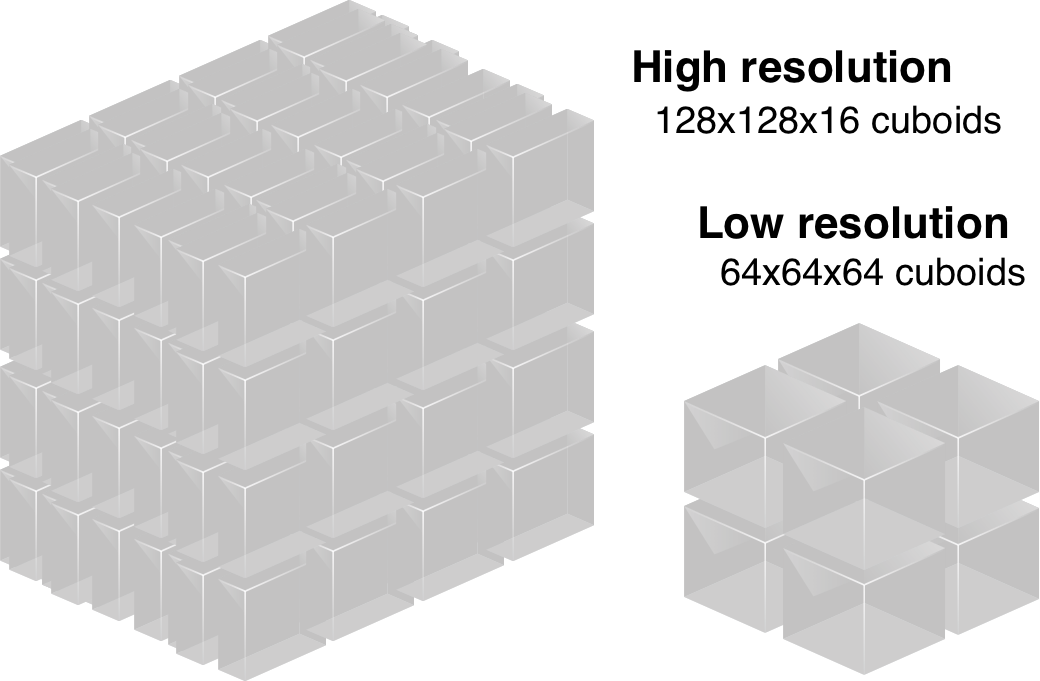}
\caption{The resolution hierarchy scales the X,Y dimensions of cuboids, but not Z.  So that cuboids contain
roughly equal lengths in all dimensions. }
\label{fig:hierh}
\end{center}
\end{figure}

Resolution scaling and workload dictate the the dimension and shape of cuboids.  
For EM data, computer vision algorithms operate on data regions that are roughly cubic 
in the original sample so that they can detect brain anatomy in all orientations. 
Because the X and Y dimensions are scaled, but not Z, this results in different sized voxel 
requests at different resolutions.  For this reason, use different shaped cuboids at different levels in 
the hierarchy (Figure \ref{fig:hierh}) 
For example, at the highest three resolutions in {\tt bock11}, cuboids are flat (128x128x16) because each voxel represents 10 times as much length in Z as X and Y.
Beyond level 4, we shift to a cube of (64x64x64).  
We include time series in the spatial index, using a 4-d space filling curve.  Time is often 
as large as other dimensions, 1000s of time points in MR data.  This supports 
queries that analyze the time history of a smaller region.  We do not include channel data in 
the index---there are different cuboids for different channels---because the number of channels tend to 
be few (up to 17) and most analyses look at combinations of a few channels.

Cuboids contain only $2^{18} = 256$K of data, which is a compromise among the different uses of the data.
This size may be too small to optimize I/O throughput.  Space filling curves mitigate this by ensuring
that larger aligned regions are stored sequentially and can be read in a single streaming I/O.
An important use of our service extracts lower-dimensional projections, for visualization of EM data and 
sub-space clustering of 4-d and 5-d data.  Keeping the cuboid size small reduces the total amount of data
to be read (and discarded) for these queries.

\subsection{Annotations}
\label{sec:anno}

An annotation project contains a description of the spatial extent and metadata for objects detected
in an image database.  Each project has a spatial database registered to an image data set.  Projects
may be shared among teams of human annotators that want a coherent view of what structures have been 
labeled.  Projects are also plentiful; each parameterization of a computer vision algorithm may have a 
project so that outputs can be compared.  Annotations also have a resolution hierarchy so that large structures, 
such as cell bodies, may be found and marked at low resolution and detail, synapses and dendritic spines, at
high resolution.

An ``annotation'' consists of an object identifier that is linked to object metadata 
in the RAMON (Reusable Annotation Markup for Open coNectomes) neuroscience ontology \cite{cajal3d} 
and a set of voxels labeled with that identifier in the spatial database.
RAMON is a hierarchical systems engineering framework for
computer-vision and brain anatomy that include concepts such as synapses, seeds, neurons, organelles, etc.
We developed RAMON for use within the Open Connectome Project community of collaborators; it is not a standard.
Although annotations are often sparse, we store them in dense cuboids.  We found that this design is 
flexible and efficient.  When annotations are dense, such as the output of the automatic image 
segmentation \cite{Reina2011a}, storing them in cuboids outperforms sparse lists by orders of 
magnitude.  To improve performance when annotations are sparse, 
we allocate cuboids lazily; 
regions with no annotations use no storage and can be ignored during reads.  We also gzip compress cube data.
Cube labels compress well because they have low entropy with both many zero values and long repeated 
runs of non-zero values (labeled regions).  Recent implementations of run length encoding \cite{Abadi2006,Wu2009}  
may be preferable, but we have not evaluated them.

We support multiple annotations per voxel through {\em exceptions}.   An exceptions list per cuboid tracks 
multiply-labeled voxels.  Exceptions are activated on a per project basis, which incurs a minor 
runtime cost to check for exceptions on every read, even if no exceptions are defined.
Each annotation write specifies how to deal with conflicting labels, by overwriting or preserving the previous label
or by creating an exception.

For performance reasons, we initially store annotations at single level in the resolution hierarchy and 
propagate them to all levels as a background, batch I/O job.  
The consequence is that annotations are only immediately visible at the resolution at which they are written.  
The alternative would be to update all levels of the hierarchy for each write.  While there are data structures and
image formats for this task, such as wavelets \cite{vapor}, 
the incremental maintenance of a resolution hierarchy always makes updates more complex and expensive. 
The decision to not make annotations consistent instantaneously reflects how we perform vision and analysis.
Multiple algorithms are run to detect different structures at different levels and fuse them together.
Then, we build the hierarchy of annotations prior to performing spatial analysis.  
Because write I/O performance limits system throughput, we sacrifice data consistency to optimize this
workflow.  We may need to revisit this design decision for other applications.

\subsection{Tiles}

At present, we store a redundant version of the image data in a 2-d tile stack as storage 
optimization for visualization in 
the Web Viewer CATMAID \cite{catmaid}.  Tiles are 256x256 to 1024x1024 image sections used in a 
pan and zoom interactive
viewer.  CATMAID dynamically loads and prefetches tiles so that the user can experience
continuous visual flow through the image region.   

The cutout service provides the capability to extract image planes as an alternative to storing tiles.
To do so, it reads 3-d cubes, extracts the requested plane, and discards the vast majority of 
the data.  To dynamically build tiles for CATMAID, we use an {\tt http} rewrite rule to convert a 
request for a file {\tt url} into an invocation of the cutout Web service.

\begin{figure}
\begin{center}
\includegraphics[width=1.4in]{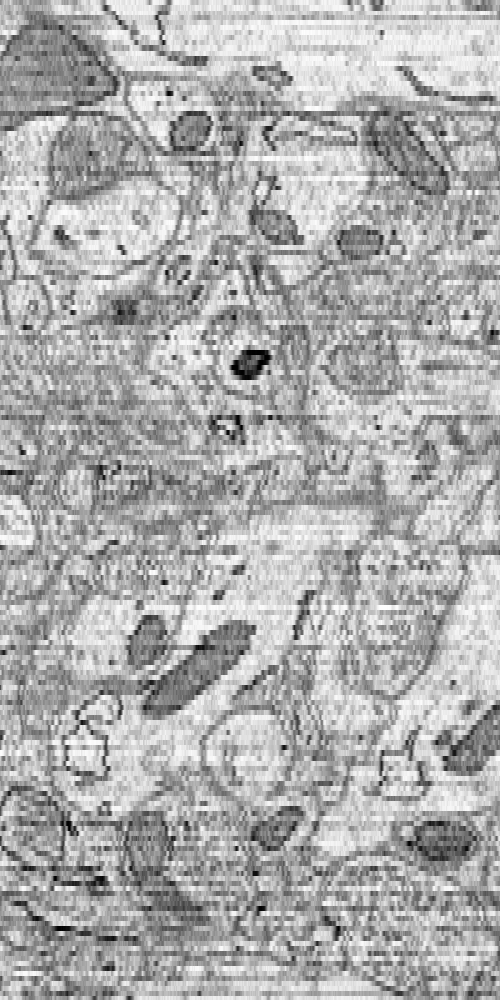}
\hspace{5pt}
\includegraphics[width=1.4in]{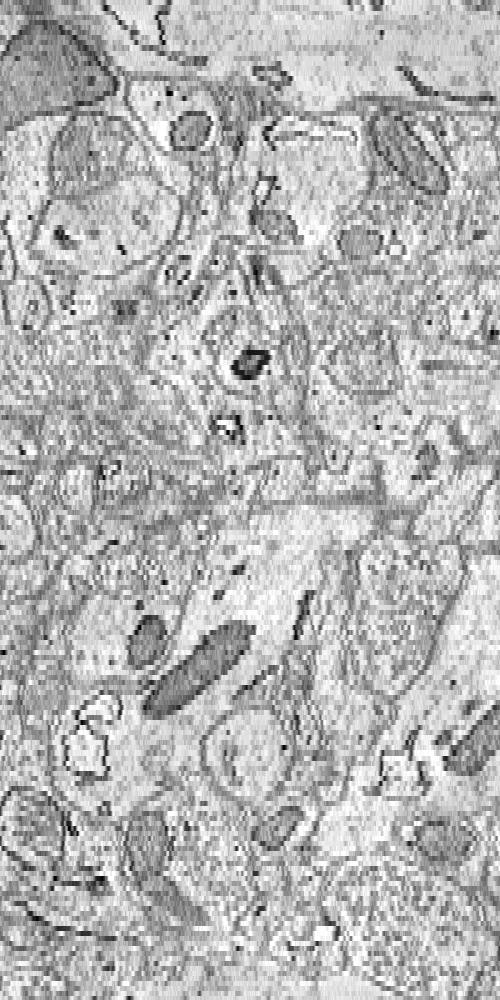}
\caption{Original (left) and color corrected (right) images across multiple serial sections \cite{kasthuri11}.}
\label{fig:xz}

\vspace{-20pt}
\end{center}
\end{figure}

\begin{figure*}[t]
\begin{center}
\includegraphics[width=6.8in]{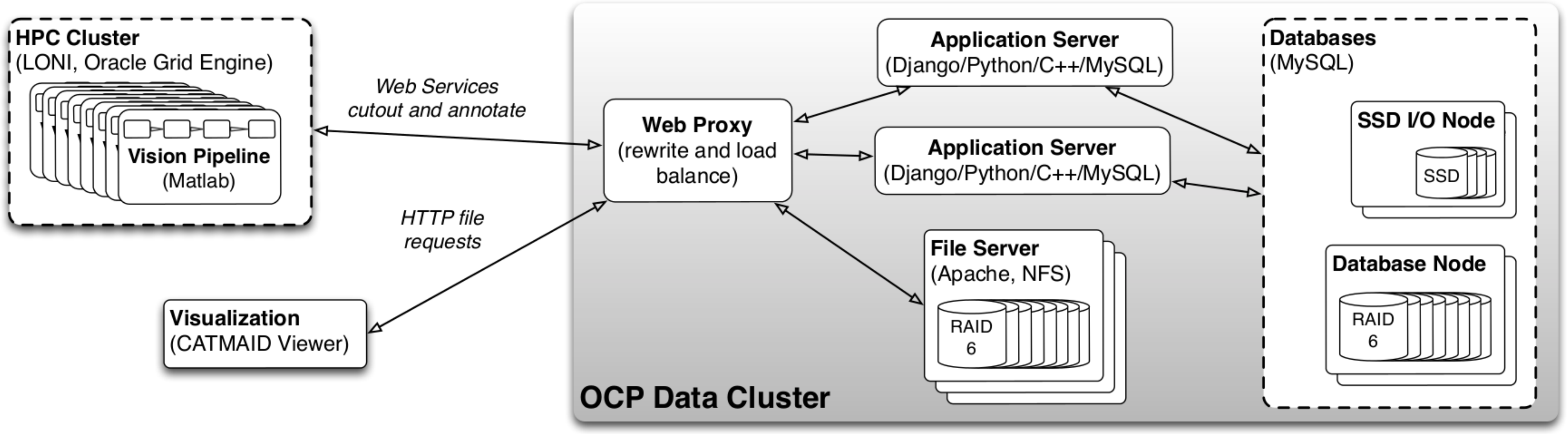}
\caption{OCP Data Cluster and clients as configured to run and visualize the parallel computer vision workflow for synapse detection (Section \ref{sec:data}).} 
\label{fig:arch}
\end{center}
\end{figure*}

Our current practice stores tiles for the image plane--the dimension of highest isotropic resolution--and 
dynamically builds tiles from the cutout service for the orthogonal dimensions including time.  
Most visualization is done in the image plane, because anisotropy, exposure differences and imperfect 
registration between slices makes images more difficult to interpret (Figure \ref{fig:xz}).  
We modify CATMAID's storage organization to create locality within each file system directory.  
By default, CATMAID keeps a directory for each slice so that a 
tile at resolution $r$ at coordinates $(x,y,z)$ is named {\tt z/y\_x\_r.png}.  This places files from 
multiple resolutions in the same directory, which decreases lookup speed and sequential I/O.  
Again using {\tt http} rewrite rules, we restructure the hierarchy into {\tt r/z/y\_x.png}.  This halves the number of
files per directory and, more importantly, each directory corresponds to a single 
CATMAID viewing plane.

We believe that we can provide performance similar to tiles in the cutout service through a combination of 
prefetching and caching.  Doing so would eliminate the redundant storage of tiles.  Instead of doing 
a planar cutout for each tile, we could round the request up to the next cuboid and materialize
and cache all the nearby tiles either on the server or in a distributed memory cache.  This is future work.

\subsection{Data Cleaning}

While not a database technology, we mention that we color correct image data to eliminate 
exposure differences between image planes.  The process solves a global Poisson equation to 
minimize steep gradients in the low frequencies, outputting smoothed data.  High-frequencies from 
the original image are added back in to preserve the edges and structures used in computer vision.
The process was derived from work by Kazhdan \cite{Kazhdan:SIGGRAPH:2008}.
The resulting color corrected data are easier to interpret in orthogonal views (Figure \ref{fig:xz}).  
We also believe that they will improve the accuracy of computer vision: a hypothesis that we are exploring.

\section{System Design}

We describe the OCP Data Cluster based on the configuration we used for running the synapse detector (Figure \ref{fig:arch}), 
because this demonstrates a parallel computer vision application.  The figure includes parallel vision pipelines
using Web services and a visualization function run from a Web browser both connecting over the public Internet.

\subsection{Architecture}
\label{sec:arch}

The OCP Data Cluster consists of heterogeneous nodes each designed for a different function or workload.
These include application servers, database nodes, file system nodes, and SSD I/O nodes.
The database nodes store all of the image and annotation data for cutout (except for high-I/O annotation 
projects on SSD I/O nodes).  
We currently have two Dell R710s with dual Intel Xeon E5630 quad-core processors, 64GB of memory,
and Dell H700 RAID controllers.  The twelve hot-plug drive bays hold a RAID-6 array of 11 2TB (resp. 3TB) SATA drives and 
a hot-spare for an 18TB (resp. 27TB) file system.  Two drives have failed and rebuilt from the hot spare 
automatically in the last 18 months.  We use MySQL for all database storage.

Application servers perform all network and data manipulation, except for database query processing.
This includes partitioning spatial data requests into multiple database queries and assembling, rewriting, 
and filtering the results.  We currently deploy two Web-servers in a load-balancing proxy.  They run on the same
physical hardware as our database nodes, because processor capabilities of these nodes exceed the compute 
demand of the databases.  These functions are entirely separable.

File server nodes store CATMAID image tiles for visualization, image data streamed from the instruments over the 
Internet, and other ``project'' data from our partners that needs to be ingested into OCP formats.  
The nodes are designed for capacity and sequential read I/O.  We have a Dell R710 with less memory (16GB) and compute
(one E5620 processor) than the database nodes.  The machine has the same I/O subsystem with a 27TB file system.
We use two Data-Scope nodes (described below) as file servers that each have a 12TB software RAID 10 array 
built on 24 1TB SATA disks.

SSD I/O nodes are designed for the random write workloads generated by parallel computer
visions algorithms. 
The two nodes are Dell R310s with an Intel Xeon 3430 processor with 4 2.4GHz cores.
Each machine has a striped (RAID 0) volume on two OCZ Vertex4 solid states drives.  The system realizes
20K IOPS of the theoretical hardware limit of 120K; an improved I/O controller is needed.
We deployed the SSD I/O nodes in response to our experience writing 19M synapses in 3 days when running 
the parallel synapse finding algorithm.  
For synapse finding, we had to throttle the write rate to 50 concurrent 
outstanding requests to avoid overloading the database nodes.  

Our growth plans involve moving more services into the Data-Scope \cite{datascope}: a 90 node data-intensive cluster 
with a 10PB ZFS file system.   We are currently using two machines as file system nodes and have an 
unused allocation of 250 TB on the shared file system.  Data-Scope nodes have 2 Intel Xeon 5690 hex core processors, 
64 GB of RAM, 3 Vertex3 SSDs and 24 local SATA disks and, thus, serve all of our cluster functions well.

\para{Data Distribution}
We place concurrent workloads on distinct nodes in order to avoid I/O interference.
This arises in two applications.  Computer vision algorithms, e.g.~our synapse detector,
reads large regions from the cutout service and performs many small writes.
We map cutouts to a database node and small writes to an SSD node.  Visualization reads image data 
from either the tile stack or cutout database and overlays annotations.  Again, annotation
databases are placed on different nodes than image databases.  Notably, the cutouts and tile
stack are not used together and could be placed on the same node.
Because SSD storage is a limited resource, OCP migrates databases from SSD nodes to database nodes
when they are no longer actively being written.  This is an administrative action implemented with 
MySQL's dump and restore utilities and most often performed when we build the annotation
resolution hierarchy (Section \ref{sec:anno}).


We shard large image data across multiple database nodes by partitioning the Morton-order
space filling curve (Figure \ref{fig:zorderpart}).  Currently, we do this only for our largest 
data set ({\tt bock11}) for capacity reasons.
We have not yet found a performance benefit from sharding
data.  For sharded databases, the vast majority of cutout requests go to a single node.
We do note that multiple concurrent users of the same data set would benefit from 
parallel access to the nodes of a sharded database.
Our sharding occurs at the application level.  The application is aware of the data distribution
and redirects requests to the node that stores the data.  
We are in the process of evaluating multiple NoSQL systems in which 
sharding and scaleout are implemented in the storage system, rather than the application.
We are also evaluating SciDB \cite{scidb,scidb2} for use as an array store for spatial data.


\subsection{Web Services}


Web-services provide rich object representations that both capture the output of computer vision algorithms and 
support the spatial analysis of brain anatomy.  We have used the service to perform
analysis tasks, such as spatial density estimation, clustering, and building distance distributions. 
Analysis queries typically involve selecting a sets of objects based on metadata properties, examining 
the spatial extent of these objects, and computing statistics on spatial properties such as distances or 
volumes.  For example, one analysis looked at the distribution of the lengths of dendritic spines, 
skinny connections from dendrites to synapses in order to understand the region of influence of neural
wiring.  The OCP queries used metadata to find all synapses of a certain type that connect to the selected
dendrite and extracted voxels for synapses and the object.  

All programming interfaces to the OCP Data Cluster use RESTful (REpresentational State Transfer)~\cite{Fielding02} interfaces
that are stateless, uniform, and cacheable.  Web-service invocations perform HTTP GET/PUT/DELETE requests to human readable
URLs.  REST's simplicity makes it easy to integrate services into many environments; we have built
applications in Java, C/C++, Python, Perl, php, and Matlab.  

We have selected HDF5 as our data interchange format.  HDF5 is a scientific data model and format.  We prefer it to 
more common Web data formats, such as JSon or XML, because of its support for multidimensional arrays and large data sets.
\begin{table*}
\small
\begin{tabular}{ | l  | l |}
\hline
3-d image cutout  & \url{http://openconnecto.me/}{\em token}\url{/hdf5/}{\em resolution/x-range/y-range/z-range/} \\
$\ldots$ \hspace{2pt} $512^3$ at offset $(1024,1024,1024)$ resolution 4 & \url{http://openconnecto.me/bock11/hdf5/4/512,1024/512,1024/512,1024/} \\
\hline
Read an annotation & \url{http:///openconnecto.me/}{\em token/identifier/data options}/ \\
$\ldots$ \hspace{2pt} and voxel list for identifier $75$ & \url{http://openconnecto.me/annoproj/75/voxels/} \\
$\ldots$ \hspace{2pt} and bounding box & \url{http://openconnecto.me/annoproj/75/boundingbox/} \\
$\ldots$ \hspace{2pt} and cutout restricted to a region & \url{http://openconnecto.me/annoproj/75/cutout/2/1000,2000/1000,2000/10,20/} \\
\hline
Write an annotation & \url{http://openconnecto.me/}{\em token/data options/} \\
\hline
Batch read & \url{http://openconnecto.me/annproj/1000,1001,1002/} \\
\hline
Predicate query (find all synapses) & \url{http://openconnecto.me/annoproj/objects/type/synapse/} \\
\hline
\end{tabular}
\normalsize
\caption{RESTful interfaces to the OCP Cutout and Annotation Web services.  Italics indicate arguments and parameters.
{\tt annoproj} is an example annotation project.}
\label{fig:urls}
\end{table*}

\para{Cutout} Providing efficient access to arbitrary sub-volumes of image data guides the design to the OCP Data System.  
The query, which we call a {\em cutout}, specifies a set of dimensions and ranges within each dimension in a URL, 
GETs the URL from the Web service, which returns an HDF5 file that contains a multidimensional array (Table \ref{fig:urls}).
Every database in OCP supports cutouts.  EM image databases return arrays of 8-bit grayscale values, annotation databases
return 32-bit annotation identifiers, and we also support 16-bit (TIFF) and 32-bit (RBGA) image formats.

\para{Projects and Datasets}  A {\em dataset} configuration describes the dimensions of spatial databases.  
This includes the number of dimensions (time series and channels), the size of each dimensions, and the 
number of resolutions in the hierarchy.  A {\em project} defines a specific database for a dataset, including 
the project type (annotations or images), the storage configuration (nodes and sharding), and properties, such as 
does the database support exceptions? and, is the database read-only? 
We often have tens of projects for a single dataset, including original data, cleaned data, and multiple
annotation databases that describe different structures or use different annotations algorithms.


\para{Object Representations}  Annotation databases combine all annotations of the same value (identifier) 
into an object associated with metadata and support spatial queries against individual objects.  The service provides
several {\em data options} that allow the query to specify whether to retrieve data and in what format (Table \ref{fig:urls}).
The default retrieves metadata only from the databases tables that implement the RAMON ontology.    
A query may request a {\tt boundingbox} around the annotation, which queries a spatial index but
does not access voxel data.  All the data may be retrieved as a list of {\tt voxels} or as a dense array by specifying 
{\tt cutout}.  The dense data set has the dimensions of the bounding box.  Dense data may also be restricted to a specific
region by specifying ranges.  Data queries retrieve voxel data from cuboids and then filter out all voxel labels that 
do not match the requested annotation.

\begin{figure}
\begin{center}
\includegraphics[width=1.6in]{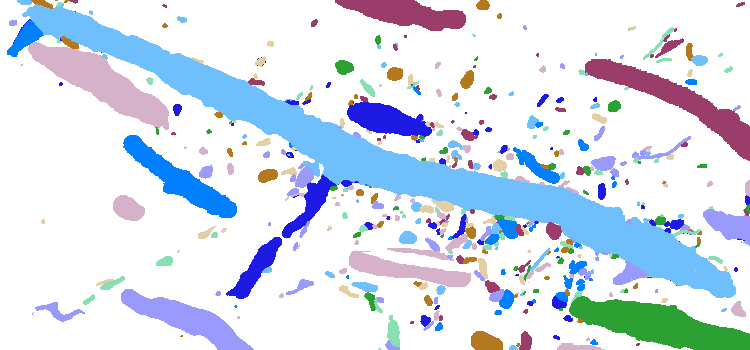}
\includegraphics[width=1.6in]{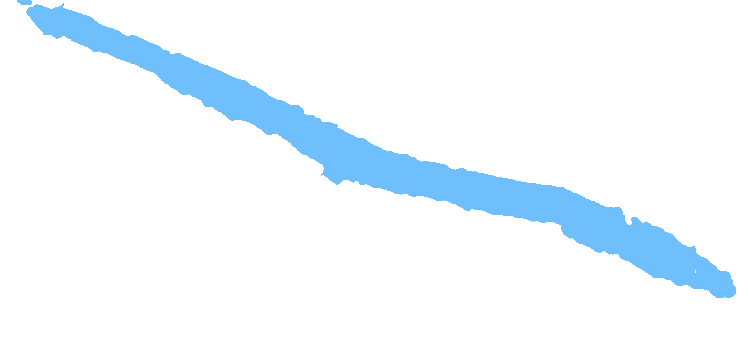}
\caption{A cutout of an annotation database (left) and the dense read of a single annotation (right).}
\vspace{-10pt}
\end{center}
\end{figure}

We provide both sparse (voxels lists) and dense (cutout) data interfaces to provide different performance options when 
retrieving spatial data for an object.  The 
best choice depends upon both the data and the network environment.  At the server, it is always faster to compute
the dense cutout.  The server reads cuboids from disk and filter the data in place in the read buffer.  
To compute voxel lists, the matching voxels locations are written to another array.  However, many neural structures
have large spatial extent and are extremely sparse.  For these, the voxel representations are much smaller.  
On WAN and Internet connections, the reduced network transfer time dominates additional compute time.
For example, the dendrite 13 in kasthuri11 set comprises 8 million voxels in a bounding box of more than 1.9 trillion voxels,
i.e. less that 0.4\% of voxels are in the annotation.  Other neural structures, such as synapses, are compact and dense 
interfaces always perform better.

To write an annotation, clients make an HTTP PUT request to a project that includes an HDF5 file.  All writes use
the same base URL, because the HDF5 file specifies the annotation identifier or gives no identifier, causing the 
server to choose a unique identifier for a new object.  The data options specify the write discipline for 
voxels in the current annotation that are already labeled in the database:
{\tt overwrite} replaces prior labels, {\tt preserve} keeps prior labels, and {\tt exception} keeps the prior label
and marks the new label as an exception.  Other data options include {\tt update} when modifying an existing 
annotation and {\tt dataonly} to write voxel labels without changing metadata.

\para{Batch Interfaces} OCP provides interfaces to read or write multiple annotations at once in order
to amortize the fixed costs of making a Web service request over multiple writes and reads.  Batching 
requests is particularly important when creating neural structures with little or no voxel data in which 
the Web service invocation dominates I/O costs to the database.  This was the cases with our synapse
finder (Section \ref{sec:data}) in which we doubled throughput by batching 40 writes at a time.
HDF5 supports a directory structure that we use to encode multiple RAMON objects, placing
each object in its own directory by annotation identifier.  

\para{Querying Metadata} OCP provides a key/value query interface to object metadata.  The
queries extract a list of objects that match a predicate against metadata.  We currently allow
equality queries against integers, enumerations, strings, and user-defined key/value pairs and 
range queries against floating points values.  Although a limited interface, these queries 
are simple and expressive for certain tasks.  The {\tt objects} Web service (Table \ref{fig:urls}) 
requests one or more metadata fields and values (or metadata field, inequality operator, 
and value for floating point fields) and the service returns a list of matching annotation identifiers.  
As an example, we use the url \url{openconnecto.me/objects/type/synapse/confidence/geq/0.99/} 
to visualize the highest confidence objects found by the synapse detection pipeline.

Although we do not currently provide full relational access to databases, we
will provide SQL access for more complex queries and move power users to SQL.  To date, Web services have
been sufficient for computer vision pipelines.  But, we see no reason to re-implement an SQL-equivalent 
parser and language through Web Services.

\para{Spatial Queries and Indexing Objects} Critical to spatial analysis of annotation databases are two queries: 
(1) what objects are in a region? and (2) what voxels comprise an object?  On these primitives, OCP applications build more complex queries, 
such as nearest neighbors, clustering, and spatial distributions.
OCP's database design is well suited to identifying the annotations in a region.  The service performs a cutout of the region 
and extracts all unique elements in that region.  In fact, the Numpy library in Python provides a built in function to 
identify unique elements that is implemented in C.  Locating the spatial extent of an object
requires an additional index.

\begin{figure}
\begin{center}
\includegraphics[width=\columnwidth]{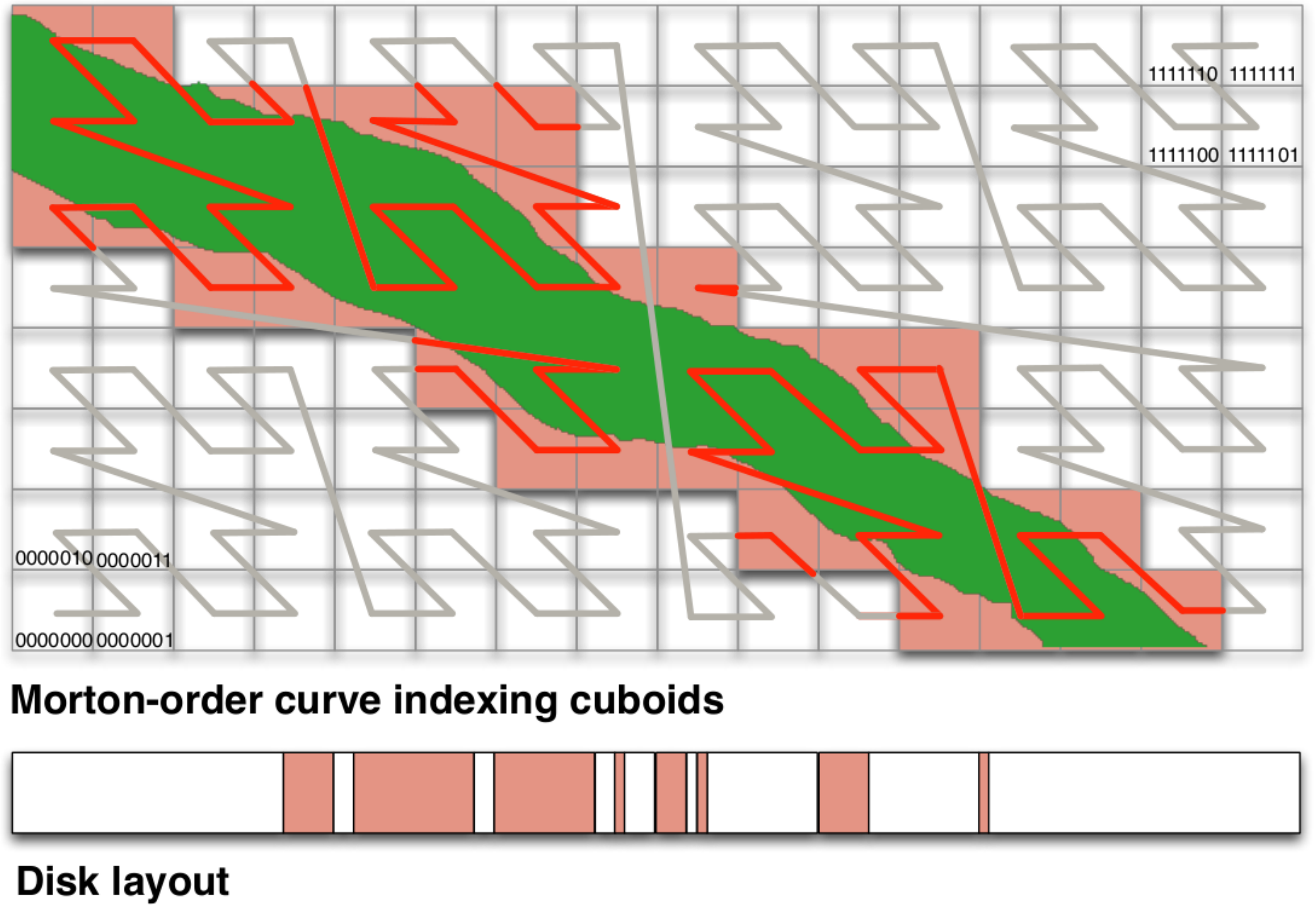}
\caption{The sparse index for an object (green) is a list of the Morton-order location of the cuboids that contain voxels for  that annotation.  The index describes the disk blocks that contain voxels for that object, which can be read in a single pass.}
\label{fig:index}
\end{center}
\end{figure}

OCP uses a simple, sparse indexing technique that supports batch I/O to locate the spatial extent of an object, its voxels,
and bounding box.  The index comprises a database table that enumerates the list of cuboids that contain voxels for each 
annotation identifier (Figure \ref{fig:index}).  The list itself is a BLOB that contains a Python array. 
Although this index is not particularly compact, it has several desirable properties.   

Adding to the index is efficient.  It is a batch operation and uses append only I/O.    
When writing an annotation in a cuboid, we determine that the annotation is new to the 
cuboid, based on the current contents, and add the cuboid's Morton-order location 
to a temporary list.  
After all cuboids have been updated, a single write transaction appends all new cuboids to the list as a batch.
The annotation process, be it manual of machine vision, tends to create new objects and only rarely deletes or prunes
existing objects.  The workload suits an append-mostly physical design.  

Retrieving an object allows for batch I/O and retrieves all data in a single sequential pass.
The query retrieves the list of cuboids and sorts then by Morton-order locations.  Then all cuboids are requested as a
single query.  Cuboids are laid out in increasing Morton order on disk and, thus, are retrieved
a single sequential pass over the data.

We chose this design over existing spatial indexing techniques because it is particularly well suited to neuroscience
objects which are sparse and have large spatial extent.  For OCP data, indexes that rely on bounding boxes 
either grow very large (R-Trees \cite{Guttman84}) or have inefficient search (R+-Trees \cite{Sellis87}).  
Informally, neural objects are long and skinny and there are many in each region so that bounding boxes intersect and 
overlap pathologically. 
An alternative is to use directed local search techniques that identify objects by a centroid and locate the object
by searching nearby data \cite{Papa06,Tauheed2012}.  For OCP data, these techniques produce much more compact indexes that are faster to 
maintain.  However, querying the index requires many I/Os to conduct an interactive search and we prefer lookup performance to 
maintenance costs.  We plan to quantify and evaluate these informal comparisons in the future.

\begin{figure*}[t]
\begin{center}
\hspace{-10pt}
\subfigure[Maximum]{
\includegraphics[height=2.0in]{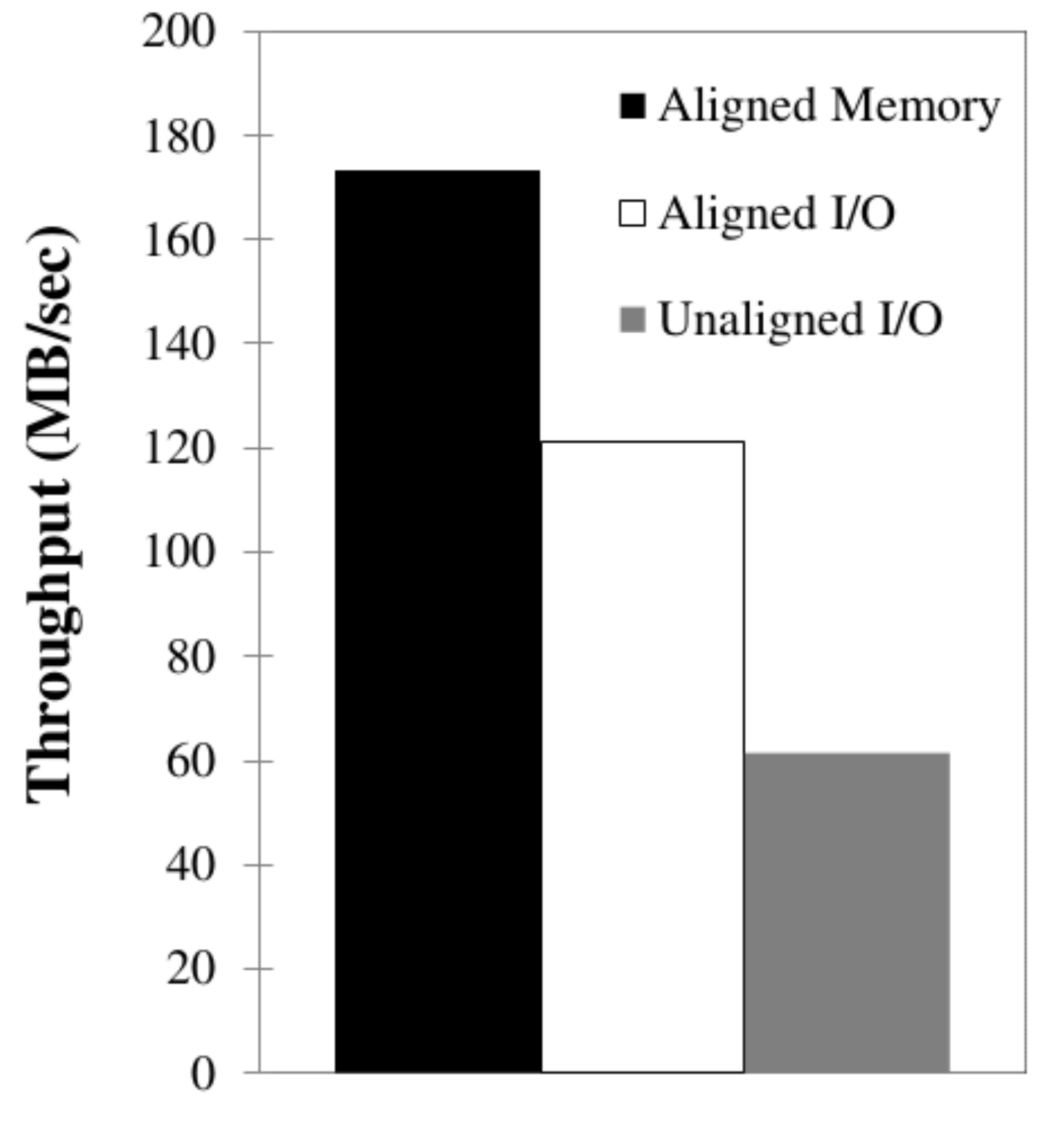}
}
\subfigure[By cutout size.]{
\includegraphics[height=2.0in]{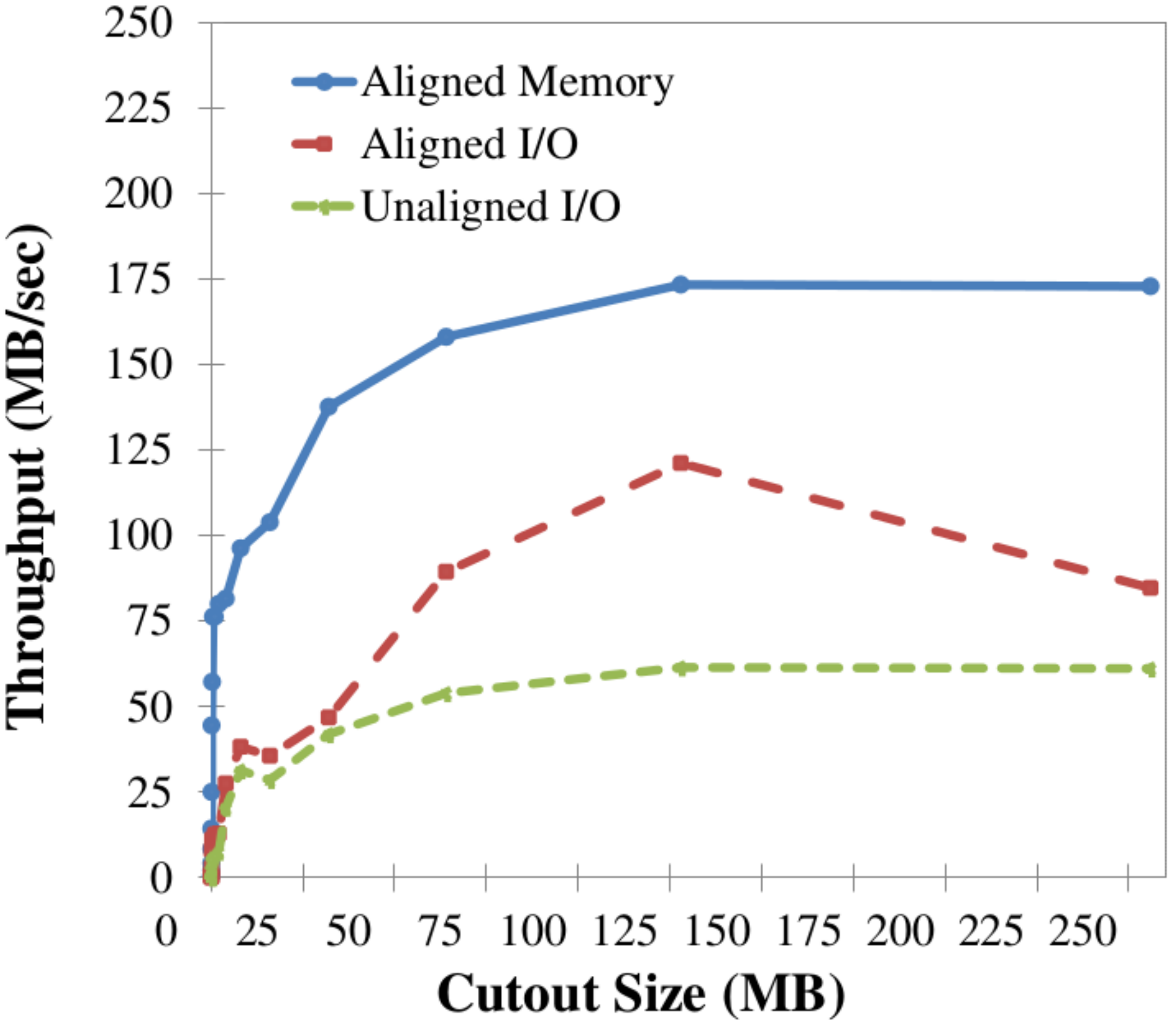}
}
\subfigure[By cutout size (log/log scale).]{
\includegraphics[height=2.0in]{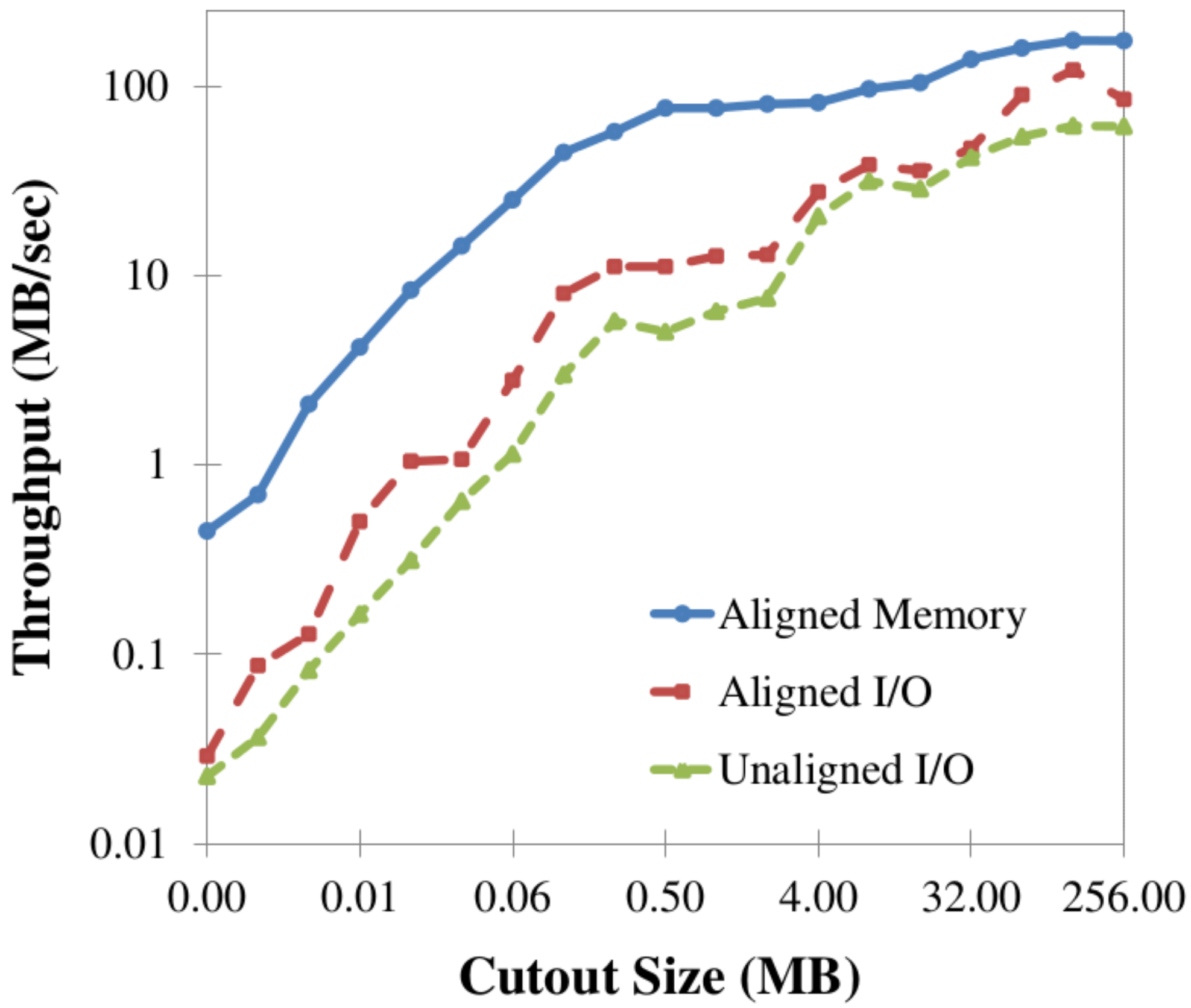}
}
\hspace{-10pt}
\caption{The performance of the cutout Web-service that extracts three-dimensional subvolumes from the {\tt kasthuri11}  image database.}
\label{fig:cutout}
\end{center}
\end{figure*}

\para{Software} Application servers run a Django/Python stack integrated with Apache2 using WSGI that 
processes Web service requests.
This includes performing database queries, assembling and filtering spatial data, and formatting data 
and metadata into and out of HDF5 files.  
We uses parallel Cython to accelerate the most compute intensive tasks.  
Cython compiles Python code into the C language so that it executes without the overhead of interpretation.  
Parallel Cython activates OpenMP multicore parallelism within Cython routines.
The compute intensive routines that we accelerate with Cython operate against every voxel in a cutout.
Examples include (1) false coloring annotations for image overlays in which every 32-bit annotation 
is mapped to an RGBA color in an output image buffer and (2) filtering out annotations that do not match a
specified criteria, e.g.~to find all synapses in a region one queries the metadata to get a list of synapse
ids and then filters all non-matching identifiers out of the cutout.  


\section{Preliminary Performance}

We conducted experiments against the live Web-services to provide an initial view of performance.
The major conclusion is that memory and I/O performance both limit throughput. 
The process of array slicing and assembly for cutout requests keeps all processors fully utilized 
reorganizing data in memory.
We had not witnessed the memory bottleneck prior to this study because OCP serves data over a 
1 Gbps Ethernet switch and a 1 Gbps Internet uplink, which we saturate long before other cluster resources.
As we migrate to the Data-Scope cluster (Section \ref{sec:arch}),  we will upgrade to a 40 Gbps Internet2 uplink and more compute capable nodes, which will make memory an even more critical resource.

Experiments were conducted on an OCP Database node: a Dell R710 with two Intel Xeon E5630 quad-core processors,
64GB of memory, and a Dell H700 RAID controller managing a RAID-6 array of 11 3TB SATA drives for a 27TB file system.  
All Web service requests are initiated on the same node and the use the localhost interface, allowing us to test the server 
beyond the 1 Gbps network limit.  Experiments show the cutout size, not the size of the data transferred or read 
from disk.  Each cuboid is compressed on disk, read, and uncompressed.  The data are then packaged into a cutout, 
which is compressed prior to data transfer.  The cutout size is the amount of data that the server must handle in memory.  
The EM mouse brain data in question has high entropy
and tends to compress by less than 10\%.  The annotation data are highly compressible.  

\begin{figure}
\begin{center}
\includegraphics[height=2in]{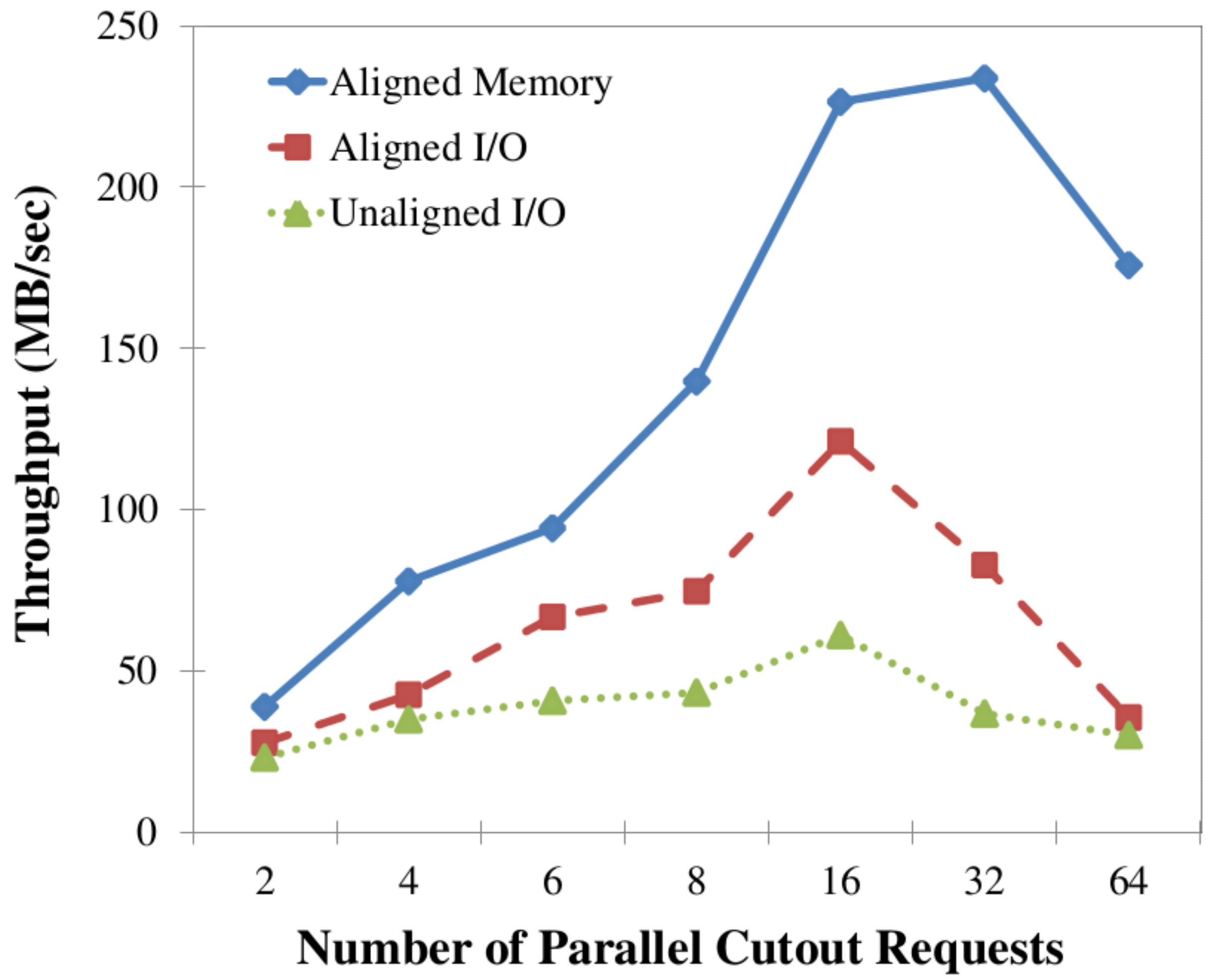}
\caption{Throughput of 256MB cutout requests to {\tt kasthuri11} as a function of the number of concurrent requests.}
\label{fig:cutout_parallel}
\end{center}
\end{figure}

Cutout throughput is the principal performance measure for our system, dictating how much data the service
provides to parallel computer vision workflows.  Figure \ref{fig:cutout}(a) shows the maximum throughput 
achieved over all configurations when issuing 16 parallel cutout requests.
When data are in memory and cutout requests are aligned to cuboid boundaries (aligned memory), 
the system performs no I/O and minimally rearranges data.  Processing in the application stack
bounds performance.   In this case, a single node realizes a peak throughput of more than 173 MB/s 
for the largest transfers.  Cutouts to random offsets aligned to cuboid boundaries add I/O costs to each request 
and bring performance down to a peak of 121 MB/s.  
Unaligned cutouts require data to be reorganized in memory, moving 
byte ranges that are unaligned with the cache hierarchy.  This incurs further performance penalties and 
peak throughput drops to only 61 MB/s.  The I/O cost of these requests is only marginally more than aligned
cutouts, rounding each dimension up to the next cuboid.  Unaligned cutouts reveals the dominance of 
memory performance.  

Scalability results reveal fixed costs in both Web-service invocation and I/O that become amortized
for larger cutouts.  Figures \ref{fig:cutout}(b) and \ref{fig:cutout}(c) show the 
throughput as a function of cutout size in normal and log/log scale.  
The experiments uses 16 parallel requests each.
Performance scales nearly linearly until 256K for reads (I/O cost) and 
almost 1 MB for in-cache (Web-service costs).  Beyond this point, performance continues to increase, albeit more slowly.
For aligned and unaligned reads, we attribute the continued increase to the Morton-order space-filling curve.  
Larger cutouts intersect larger aligned regions of the Morton-order curve producing larger contiguous I/Os \cite{moon01}.
We do not report performance above a 256M cutout size.  Beyond this point, the buffers needed by the application and
Web server exceed memory capacity.

To realize peak throughput, we had to initiate multiple requests in parallel.  The application stack runs each 
Web-service request on a single process thread and memory limits the per-thread performance.  Figure 
\ref{fig:cutout_parallel} shows throughput as a function of the number of parallel requests.
Throughput scales with the number of parallel requests beyond the eight physical cores of the machine to 16 when reading data from 
disk and to 32 when reading from memory. Performance scales beyond the 8 cores 
owing to some combination of overlapping I/O with computation and hyperthreading.  Too much parallelism eventually 
leads to a reduction in throughput, because all hardware resources are fully utilized and more parallelism 
introduces more resource contention.

Parallelism is key to performance in the OCP Data Cluster and the easiest way to realize it is through initiating 
concurrent Web-service requests.  Parallelizing individual cutouts would benefit the performance of individual 
request, but not overall system throughput, because multiple requests already consume the memory resources of all 
processors.
We note that the Numpy Python library we use for array manipulation implements parallel operators using 
BLAS, but does not parallelize array slicing.

\begin{figure}
\begin{center}
\includegraphics[height=2in]{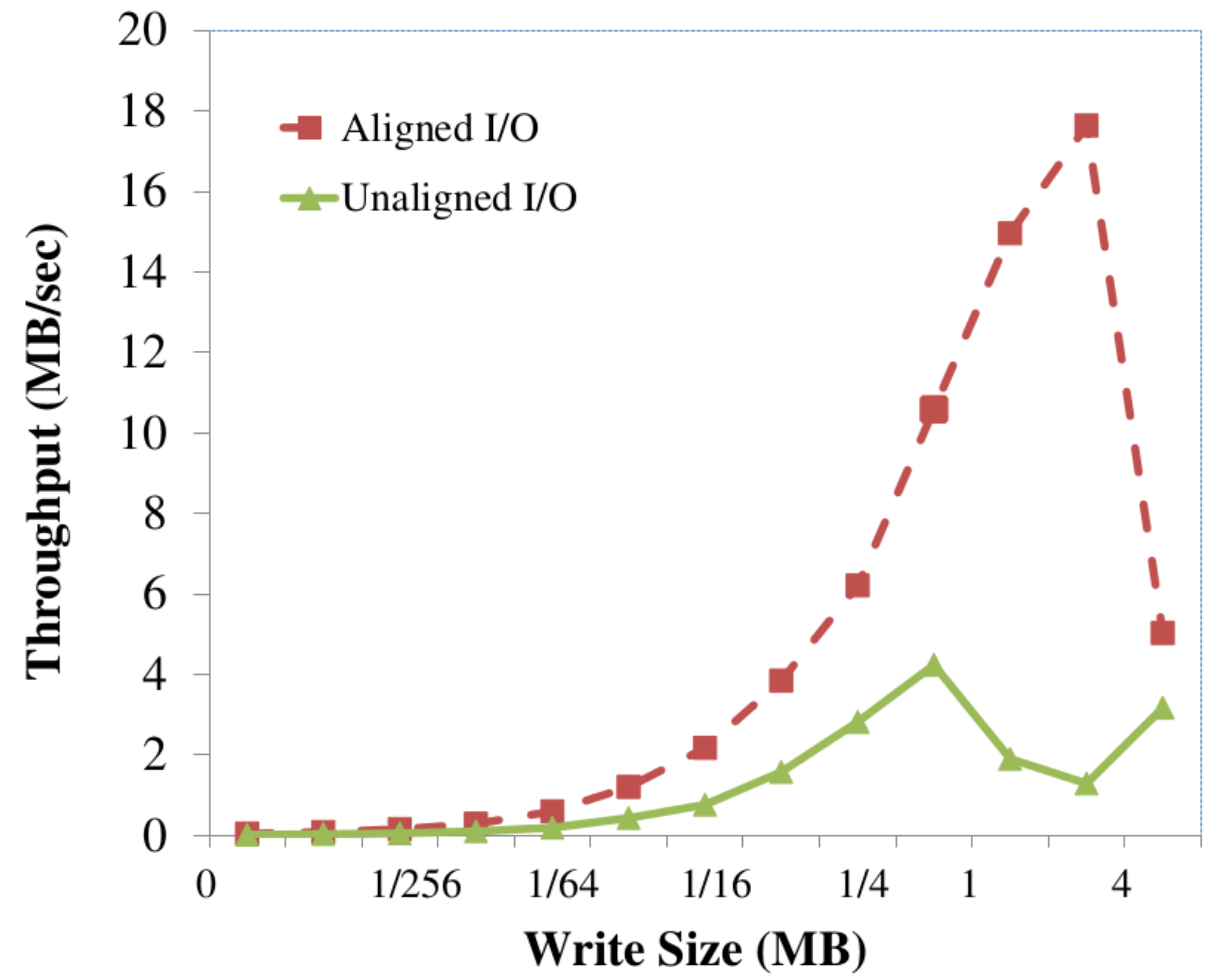}
\caption{Throughput of writing annotations as a function of the size of the annotated region.}
\label{fig:write}
\end{center}
\end{figure}

Figure \ref{fig:write} shows the write throughput as a function of the volume size when uploading annotated volumes.
Annotations differ from cutouts in that the data are 32-bits per voxel and 
are highly compressible. 
For this experiment, we uploaded a region dense manual annotations of the {\tt kasthuri11} data
set in which more than 90\% of voxels are labeled.  Data compress to 6\% the original size.
However, performance scales with the uncompressed size of the region, because that dictates how many cuboids
must be manipulated in memory.  
This experiment uses 16 parallel requests.
Write performance scales well up to 2MB cutouts and is faster than read at the same cutout size,
because the data compress much better.
However, beyond 2MB transfers, write throughput collapses.  
The best performance over all configurations of 19 MB/s does not compare well with the 121 MB/s when 
reading image cutouts.
Updating a volume of annotations is much more complex than a cutout.   It (1) reads the previous annotations,
(2) applies the new annotations to the volume database, resolving conflicts on a per voxel basis, (3) writes 
back the volume database, (4) reads index entries for all new annotations, (5) updates each list by unioning new and
old cuboid locations, and (5) writes back the index.  I/O is doubled and index maintenance costs
are added on top.  
The critical performance issues is updating the spatial index. 
Parallel writes to the spatial index result in transaction retries and timeouts in MySQL due to 
contention.  
Often, a single annotation volume will result in the update of hundreds of index entries, one for
each unique annotation identifier in the volume. 
We can scale to larger writes at the expense of using fewer parallel writers.
Improving annotation throughput is a high priority for the OCP Data Cluster.    

\begin{figure}
\begin{center}
\includegraphics[width=3in]{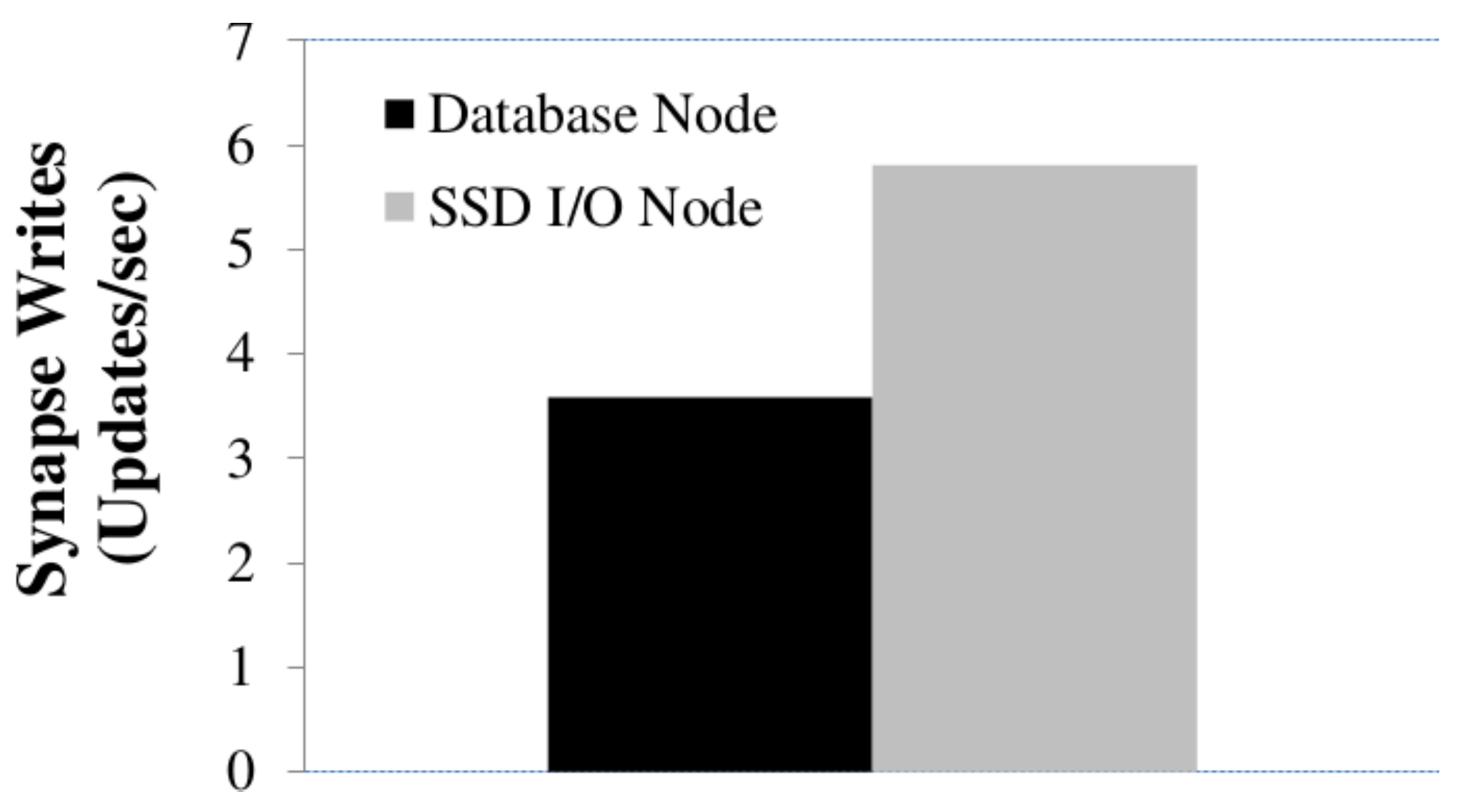}
\caption{Performance comparison of Database nodes and SSD nodes when writing synapses (small random writes).}
\label{fig:ssd}
\end{center}
\end{figure}

We also include an comparison of small write performance between an SSD node and a Database node.  The SSD node
is a Dell R310 with an Intel Xeon 3430 with 4 2.4GHz cores, 16 GB of memory, and two OCZ Vertex 4 drives configured
in a RAID 0 array.  The experiment uploads all of the synapse annotations in the {\tt kasthuri11} data in random 
order, committing after each write.  Figure \ref{fig:ssd} shows that the SSD node achieves more than 150\% the throughput of
the Database array on small random writes.  The number of synapse writes per second is surprisingly low.
We write only about 6 RAMON objects per second.  However, a single Ramon object write generates updates to 
three different metadata tables, the index, and the volume database.  
The takeaway from this experiment is that an inexpensive SSD node (<\$3000) 
offloads the write workload of an entire Database node (>\$18,000).  All I/O in this experiment is random. 
In practice, we achieve much higher write throughput because of locality and batching of requests.  Our synapse finder
workload uploaded more than 73 synapses a second per node, but that number reflects caching and effects when 
updating many synapses in a small region.

\section{Related Work}

The performance issues faced by the spatial databases in the OCP Data Cluster parallel those of SciDB \cite{scidb,scidb2},  
including array slicing, selection queries, joins, and clustering/scale-out.  SciDB might benefit OCP
by making data distribution transparent and by offloading application functions implemented in Python 
to SciDB. 

Many of the physical design principles of OCP follow the ArrayStore of Soroush \cite{Soroush11}.
They give a model for multi-level partitioning (chunking and tiling) that is more general than OCP cuboids.
They can handle both regular and irregular data. 
ArrayStore extends decades of work on regular \cite{Chang97,Moon98} and irregular \cite{Chang98} tiling.
The RasDaMan multidimensional array database \cite{Baumann98} also builds spatial data services
for image processing on top of a relational database. 
They too have explored the issues of tiling, data organization, and 
data distribution \cite{Furtado99}.   In contrast to general-purpose array 
databases, OCP has a focus on the specific application of parallel computer vision for neuroscience and, as such, 
tunes its design to application concepts and properties.  Examples include using the partitioning of 
space-filling curves in the data distribution function and indexing of neural objects.

OCP's design incorporates many techniques developed in the spatial data management community.  Samet
\cite{samet06} wrote the authoritative text on the subject, which we have used as a reference for 
region quad-trees, space-filling curves, tessellations, and much more.

OCP represents the a current state of evolution of the scale-out database architectures developed by the 
Institute for Data-Intensive Engineering and Science at Johns Hopkins.  Gray and Szalay developed
the progenitor system in the Sloan Digital Sky Survey~\cite{sdss}.  This has lead to more than 30 different 
data products over the past decade, including the JHU Turbulence Database Cluster \cite{Li08} and 
the Life Under Your Feet soil ecology portal \cite{luyf}.

\section{Final Thoughts}

In less than two years, The Open Connectome Project has evolved from a problem statement to a
data management and analysis platform for a community of neuroscientists with the singular 
goal of mapping the brain.  
High-throughput imaging instruments have driven data management off the 
workstation and out of the lab into data-intensive clusters and Web-services.
Also, the scope of the problem has forced experimental biologists to engage statisticians, 
systems engineers, machine learners, and big-data scientists.   The Open Connectome Project
aims to be the forum in which they all meet.

The OCP Data Cluster has built a rich set of features to meet the analysis needs of its users.
Application-specific capabilities have been the development focus to date.   The platform manages
data sets from many imaging modalities.  (Five to fifteen depending on how one counts them.)

Presently, our focus must change to throughput and scalability.  The near future holds two 100 
teravoxel data sets---each 
larger than the sum of all our data.  We will move to a new home in the Data-Scope cluster,
which will increase our outward-facing Internet bandwidth by a factor of forty.
We also expect the Connectomics community to expand rapidly.  The US President's Office recently 
announced a decade-long initiative to build a comprehensive map of the human brain \cite{Markoff2013}.  
This study reveals many limitations of the OCP data cluster and provides guidance to 
focus our efforts, specifically toward servicing many small writes to record brain structures
and efficient and parallel manipulation of data to alleviate memory bottlenecks.

\section*{Acknowledgments} The authors would like to thank additional members of the Open Connectome Project
that contributed to this work, including Disa Mhembere and Ayushi Sinha.  We would also like to thank 
our collaborators Forrest Collman, Alberto Cardona, Cameron Craddock, Michael Milham, Partha Mitra, and Sebastian Seung.  

\balance

\bibliographystyle{abbrv}
\bibliography{ssdbm13}

\end{document}